# Decoding Driver Takeover Behaviour in Conditional Automation with Immersive Virtual Reality

Muhammad Sajjad Ansar[a], Bilal Farooq[1a]

[a]*Laboratory of Innovations in Transportation (LiTrans), Toronto Metropolitan University, ON M5B 2K3, Canada*

**Abstract**

The safe transition from conditional automation to manual driving control is significantly intertwined with the vehicle's lateral and longitudinal dynamics. The transition may occur as a result of a system-initiated mandatory takeover (MTOR) or as a driver-initiated discretionary takeover (DTOR). In either condition, the takeover process entails differing cognitive demands and may affect the driving behaviour differently. This study analyzes driving stability and perceived mental workload in 304 takeover attempts recorded from 104 participants within virtual and immersive reality environments. Adopting an exploratory approach, this dynamic simulator study employs a mixed factorial design. Utilizing a deep neural network based survival analysis with SHAP interpretability, the study investigated the influence of covariates on perception-reaction time (PRT), distinguishing between safe and unsafe control transition and offering insights into the temporal dynamics of these shifts. The distributions of key parameters in experimental groups were analyzed and factors influencing the perceived mental workload were estimated using multivariate linear regression. The findings indicate a notable decrease in the risk of unsafe takeovers (described by a longer PRT) when drivers have prior control-transition experience and familiarity with Automated Vehicles (AVs). However, drivers' prior familiarity and experience with AVs only decreased the perceived mental workload associated with DTOR, with an insignificant impact on the cognitive demand of MTOR. Furthermore, multitasking during automated driving significantly elevated the cognitive demand linked to DTOR and leads to longer PRT in MTOR situations. The results highlight the need for extended research on takeover conditions where the decision-making process for driving control transitions is somehow voluntary, offering a subtle perspective compared to the prevailing focus on MTOR transitions in the existing literature.

## 1. Introduction

Automated vehicle technologies offer a promising future with an increased level of road safety and convenience in mobility. While integrating automated features undoubtedly enhances safety, it is imperative to acknowledge that in the short- and medium-term the automated technologies alone are unable to independently manage a vehicle on the roads. Drivers must remain consistently engaged, actively switching control with the automation system to ensure a safe driving experience. In such conditions, especially when navigating through various control transitions, the significance of maintaining stable driving is further underscored.

In conditional automation (SAE, 2021), if the driving system exits the defined operational design domains (ODD) and detects a functional limit, it immediately alerts the drivers through visual-auditory signals to regain control, providing sufficient leeway. For instance, a request-to-intervene (R2I) will be issued in the automated driving condition when the participants encounter a broken-down car in front. Subsequently, the drivers must promptly discontinue their monitoring role and safely take back control to manually change lanes to avoid the disabled vehicle. However, other motivations for switching from automated driving control can also exist when a driver takes back the control for personal comfort or prefers active involvement (Wu et al., 2022). The takeover conditions can be categorized into two distinct categories: mandatory takeover (MTOR), where drivers are compelled to intervene due to ODD limitations identified by the system and discretionary takeover (DTOR), exemplified by drivers' choice to initiate transferring the control voluntarily. While acknowledging the possibility of mandatory conditions for driver-initiated control switching (Hu et al., 2023), we specifically focus on exploring discretionary takeover scenarios in the underlying controlled experiment.

A driver's intervention ability in DTOR remains highly important despite the potential for moderate mental workload and smoother takeovers, as it reflects a proactive disengagement and

*Email addresses:* mansar@torontomu.ca (Muhammad Sajjad Ansar), bilal.farooq@torontomu.ca [1- corresponding author] (Bilal Farooq)

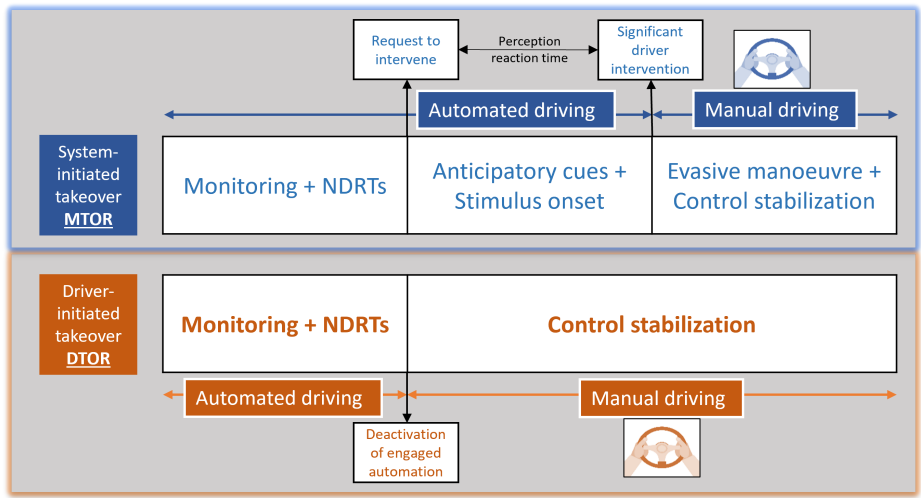

Figure 1: Vehicle control transition in MTOR and DTOR

deliberate decision-making by the drivers in managing the transition. Drivers require considerable moments to stabilize the longitudinal and lateral control of the vehicle, whether the transition to the manual is predictable or not. In cases where the automation system requests driver intervention, drivers must quickly comprehend the system's alert and make reactive decisions. This process involves rapidly navigating through information, commonly referred to as Perception-Reaction Time (PRT) in the literature (Rodak et al., 2021). Despite providing adequate audiovisual alerts by the connected system, PRT can vary based on engagement in different non-driving related tasks (NDRTs). Figure 1 illustrates the vehicle control transition process in MTOR and DTOR, progressing from left to right.

The study presents a comprehensive examination of post-takeover vehicle stability, perceived mental workload during takeovers, perception-reaction time in MTOR, and the learning effects observed across multiple discretionary takeovers of a conditionally automated vehicle. The extensive set of explanatory variables encompassed kinematic parameters, stability indices, perceptual variables (such as locus of control), attitudinal variables, socio-demographics, and environmental variables. The data collection methodology employed advanced technologies, including virtual immersive reality and digital twins, establishing a controlled and realistic experimental environment. The key contributions of this research are threefold. First, it pioneers a comprehensive analysis and modelling of both discretionary and mandatory takeover quality in conditional au-tomation, providing valuable insights into their distinct qualities. Second, the modelling results offer practical utility for road safety standard makers and automobile manufacturers, particularly in understanding PRT time-to-event dynamics and cognitive demand. Lastly, the exploration of learning effects observed in discretionary takeovers adds a unique dimension to the existing literature, contributing to our understanding of drivers' adaptive behaviour. Furthermore, the present study examines the following key research questions:

1. How does the perceived mental workload vary between discretionary and mandatory takeover situations?
2. How do drivers adapt and learn from successive discretionary takeover experiences?
3. What factors contribute to the perception-reaction time and influence a cognitive demand in different settings?

The paper progresses as follows. In Section 2, we delve into the existing literature concerning the examination of cognitive demand and driving stability across various takeover situations. Section 3 provides a comprehensive overview of the methodology employed in this study. Moving forward, Section 4 delves into the study design and expounds upon the gathered data. The model findings and their implications are discussed in Section 5. Finally, Section 6 brings the paper to a close with a concluding discussion.

## 2. Background

Disengaging the automated features in L3 driving results in the driver becoming fully responsible for operational, tactical, and strategic vehicle control. Given the multifaceted dimensions associated with the decision to transition control, the literature on takeover performance primarily explores the efficiency and safety with which the driver regains control and responds to anticipatory cues and stimulus signals. The existing evaluation encompasses NDRTs influencing takeover performance and back-to-the-loop manual control stabilization. It extends to inform the vehicle's design by exploring Human-Machine Interface (HMI) features that can improve driver takeover performance during control transitions. While existing studies predominantly address system-initiated takeovers (MTOR), there remains a notable gap in exploring driver-initiated takeovers (DTOR) and the potential influence of secondary tasks during automated driving on driver intervention when control is initiated without a system request. The following review delves into the existing literature, specifically focusing on cognitive demand and driving stability in both types of takeovers.

*2.1. Cognitive demand in MTOR and DTOR*

Studies investigating cognitive demand associated with mandatory takeovers (MTOR) are plentiful (Zhang et al., 2023; Ma et al., 2020; Choi et al., 2020). Several factors influence the cognitive demand experienced by drivers during such control transitions. These encompass the engagement in NDRTs, the driver's involvement in anticipating cues, and active communication with the vehicle's HMI. As the driving scenario becomes more intricate, involving elements like evasive manoeuvres, challenging environmental conditions, or complex traffic scenarios, the significance of cognitive demand escalates.

To measure the cognitive demand in takeovers, we recognize that three methods are frequently used in the literature. These include physiological measures like heart rate variability and pupillometry (Meteier et al., 2021; Radhakrishnan et al., 2023), task performance measures such as Detection-Response Task (Müller et al., 2021), and subjective self-assessment tools like the NASA Task Load Index (NASA-TLX) (Wei et al., 2023). Among the listed methods, the NASA-TLX perceived workload is particularly significant due to its multidimensional nature, capturing various aspects of workload like mental, physical, and temporal demands, aiding in a comprehensive cognitive load assessment (Hart and Staveland, 1988).

Drivers engage with NDRTs whether an explicit request triggers the takeover or when they switch control just based on trust and comfort. Various NDRTs differ in cognitive demands and affect drivers' ability to respond to driving-related situations. For instance, Choi et al. (2020) investigated the effects of two cognitive tasks, the Surrogate Reference Task (SuRT) and the N-back task, on takeover performance. Both tasks induced significant cognitive and visual loads, but their effects on takeover time and steering angle variance differed over time during automated driving scenarios. Furthermore, Ma et al. (2020) analyzed five workload levels defined on a controlled modality using the Tetris game, revealing an upside-down U-shaped relationship with takeover performance, aligning with the Yerkes–Dodson Law (Yerkes et al., 1908). Their findings, based on self-reported subjective perceptions (NASA TLX), indicated that a moderate workload level induced by NDRTs led to improved takeover performance, while overload or underload resulted in performance degradation. Moreover, Wandtner et al. (2018) investigated takeover performance based on reaction timing in different stimulus and response modalities. Their results revealed that visual-manual tasks (e.g., writing text messages in mounted and handheld conditions) significantly degraded takeover performance compared to tasks involving auditory-vocal and visual-vocal interactions. In general, the literature reveals clear gaps in examining cognitive models and evaluating NDRTs in the inevitable discretionary takeover scenarios.

*2.2. Driving stability in MTOR and DTOR*

Post-transition (manual) driving stability is the primary factor in evaluating the transition performance. In the review of takeover performance indicators, Hu et al. (2023) highlighted that for an efficient takeover, the speed of taking over the automation (takeover time) is as important as the driving stability after drivers take over the vehicle (takeover quality). Drivers face a significant challenge in stabilizing their vehicle's lateral and longitudinal control, regardless of whether the transition to manual control is mandatory or discretionary. For evaluating takeover quality, satisfactory or unsatisfactory thresholds have been established based on vehicle stability characterization parameters such as lateral and longitudinal acceleration, lane departure, and steering speed. For example, Radlmayr et al. (2014) refers to the takeover quality as unsatisfactory if the composite value of the vehicle's maximum longitudinal and lateral acceleration surpasses 11 m/s² during the takeover process.

The primary objective measures for assessing takeover quality are often time-based driver reaction metrics, such as driver intervention time, control stabilization time, remaining action time, and other similar parameters outlined in the driver intervention performance assessment framework by Rodak et al. (2021). However, these reaction measures may become less relevant in the context of discretionary takeovers, where the reference interval of the system request to intervene (R2I) is not applicable. This emphasizes the significance of behavioural-related vehicle dynamics, including lateral and longitudinal acceleration, rate of change of acceleration (jerk), etc., as these factors become the main determinants of takeover quality in DTOR. It underlines the need for a broader set of kinematic metrics. Additionally, relative to DTOR, driving stability becomes more critical in MTOR as the process involves evasive manoeuvre besides manual driving control stabilization (Hu et al., 2023).

The distinctive nature of drivers' situational awareness in the underlying takeover situations is another reason for different driving stability (Kamezaki et al., 2020). When drivers independently switch between manual and automated driving modes, situational awareness involves their comprehensive understanding of the driving context, road conditions, traffic patterns, and comfort level. Drivers have to assess the appropriateness of automation for the present situation, considering factors such as weather conditions, road types, traffic density, and their readiness to assume control. However, when the automation system prompts driver intervention, situational awareness extends to the driver's ability to comprehend the system's alert rapidly, identify the reason for the intervention request, and evaluate the situation's urgency. It requires swift processing of information related to the environment and the potential risks that triggered the system's request.

*2.3. Remarks*

While extensive research has been conducted on takeover quality and performance, the dynamics of DTORs have received relatively less attention in the literature. Addressing this gap is crucial for understanding the factors that influence overall driver takeover behaviour in automated vehicles. This study endeavours to bridge this gap by comprehensively analyzing cognitive demand and driving stability in both takeover conditions. By exploring how different settings influence driver intervention during each takeover event, the study sheds light on pivotal yet understudied aspects of control transitions.

**3. Methodology**

This section elaborates on the methodology for deriving lateral and longitudinal stability indices from kinematic data. We detail the process of acquiring perception-reaction time (PRT). Furthermore, we explain the formulations of applied models, covering survival analysis and multivariate linear regression.

## 3.1. Lateral and longitudinal stability

To analyze the vehicle dynamics, the average and standard deviation (std) of kinematic parameters were computed over a specific 4-second timeframe immediately following the transition. It is a relevant and significant portion of the post-transition phase. The selection of a 4-second timeframe is strategically aligned with the Time to Collision (TTC) concept in auto-pilot mode (Zhang et al., 2023), which estimates the time required for a vehicle to reach a potential collision point. This timeframe ensures that the analysis captures key dynamics and variations, contributing to the consistency and relevance of the findings within the study context.

Assessing a vehicle's lateral and longitudinal displacement involves the consideration of the magnitudes of lateral and longitudinal stability. Equation (1) delineates the calculation of lateral stability, measuring the lateral deviation derived from the trajectory data's lateral coordinates. Concurrently, equation (2) characterizes the magnitude of longitudinal stability by evaluating acceleration fluctuations exhibited by the vehicle during the driving control switching process. Higher magnitudes in these equations signify elevated levels of both lateral and longitudinal instability (Chen et al., 2023), providing valuable insights into the dynamic aspects of driving control transitions.

$$\text{lateral stability index} = \frac{\text{std}(D_x(t-40), \ldots, D_x(t))}{\text{mean}(D_x(t-40), \ldots, D_x(t))} \quad (1)$$

Where:

$D_x(t)$ represents the absolute difference in lateral coordinates between two consecutive frames, calculated as $|X(t) - X(t-1)|$

$X(t)$ denotes the lateral coordinate at frame t

$D_x(t)$ signifies the lateral offset within the time interval $[t-1, t]$

The coefficient of variation (CV) is defined as the ratio of the standard deviation (*std*) to the mean value.

$$\text{longitudinal stability index} = \frac{\sum_{i=t}^{T} |a_i - \bar{a}|}{T} \quad (2)$$

Where:

$a_i$ represents the acceleration at frame i

$\bar{a}$ denotes the average acceleration over the time period T (i.e., 4 seconds)

*3.2. Perception-Reaction Time (PRT)*

A detection response task (ISO17488:2016) is used to record the perception-reaction time (PRT) in mandatory takeover conditions. It includes both the perception phase and the subsequent reaction phase. The perception phase involves becoming aware of the external event (i.e., accident ahead) or audiovisual cues before responding. The reaction phase entails the response action, such as pulling the trigger behind the steering wheel to shift control, initiated by the human driver in response to the perceived stimulus. Together, the perception and reaction phases contribute to the total time it takes for an individual to detect and respond to a stimulus. This PRT is considered anticipated, given that drivers were informed about the experimental procedure and the required action. The average anticipated PRT is determined to be 2.63 seconds, based on the collected data.

However, for AV technology, it is essential to identify unexpected roadside conditions and promptly notify the driver within a safe Stopping Sight Distance (SSD) range. This ensures the driver has enough time to re-engage and execute the necessary evasive manoeuvre. Consequently, adjusting the obtained values to derive the surprised/unexpected PRT is necessary, aligning with established standards such as CEDR and AASHTO. To obtain the unexpected PRT, we adopted the identical correction factor (i.e., 1.35), employed by Demicoli et al. (2021), when determining the response time for L3 automated vehicles coherent with the road design standards. The calculated average for the unexpected PRT is thus 3.55 seconds. We regard the average unexpected PRT value as the safe takeover threshold, with values exceeding this threshold deemed unsafe. It is essential to highlight that tasks leading to crashes are excluded from the modelling process of safe and unsafe PRTs.

*3.3. Survival analysis of PRT*

The survival analysis technique is used to analyze the safe and unsafe PRTs. It is widely employed in medical research, epidemiology, and various other fields to study the time elapsed until a particular outcome, such as death, disease recurrence, or failure, occurs. It considers the duration each individual in a study experience until the event's occurrence or until they are censored, meaning the event has not transpired by the study's endpoint. It models the hazard function to understand how the risk of the event changes over time. Among different methods for estimating hazard function, the importance of the Cox Proportional Hazard (CPH) Model (Cox, 1972) lies in its ability to provide valuable insights into the factors influencing the timing of an event. Its primary functions include assessing the relationship between one or more predictor variables and the hazard (risk) of an event happening over time. The hazard function in the CPH model is generally represented as follows:

$$h(t|\mathbf{x}) = h_0(t) \cdot e^{\beta_1 x_1 + \beta_2 x_2 + ... + \beta_k x_k} \tag{3}$$

Where:

$h(t|\mathbf{x})$ denotes the hazard function at time t considering the covariates x,

$h_0(t)$ represents the baseline hazard function,

$\beta_1, \beta_2, .., \beta_k$ stand for the coefficients for the predictor variables $x_1, x_2, .., x_k$, and

$e^{\beta_1 x_1 + \beta_2 x_2 + ... + \beta_k x_k}$ represents the partial hazard function, indicating how the hazard changes with a unit change in the predictor variables. This function is considered independent of time.

Estimating the coefficients in the CPH model involves maximizing the partial likelihood, a statistical approach to estimating values that best align with observed survival data. This method allows for capturing the impact of each covariate on the hazard function, offering valuable insights into the factors influencing the time to an event of interest. Before conducting the estimation using the CPH model, a preprocessing step was undertaken to rank the importance of features. This ranking was based on the Relief family of algorithms, as proposed by (Robnik-Šikonja et al., 1997), with the objective of selecting relevant features to prevent overfitting and improve model performance. Following this approach, the top 19 uncorrelated features were identified for inclusion in the subsequent CPH analysis. Consequently, the CPH model was constructed based on these selected covariates, resulting in a model with relatively good predictive ability, effectively capturing the underlying dynamics of the survival data. Specifically, the final CPH model incorporated 7 covariates deemed most influential in predicting the safe and unsafe reaction times.

*3.4. Deep neural network based survival analysis of PRT*

The significance of data-driven survival analysis has grown substantially, owing to the rapid accumulation of complex and high-dimensional datasets across diverse domains. By employing data-driven methodologies, researchers are empowered to utilize advanced techniques like deep learning, extracting invaluable insights into the intricacies of survival dynamics and enhancing predictive accuracy. In this study, the deep CPH model extends the traditional CPH framework by replacing the linear log-partial hazard function with a deep neural network. The adapted hazard function formulation is expressed as follows:

$$h(t|\mathbf{x}) = h_0(t) \cdot e^{(g_w(Z_n))} \qquad (4)$$

Here, "$g_w$" represents the specified neural network architecture, characterized by its weights denoted by "$w$" The top "$n$" covariates ($Z_n$), determined by their feature importance weights, are seamlessly integrated into the network's input layer. Various architectural parameters, such as the number of covariates, hidden layers, nodes within each layer, and dropout rate, are meticulously fine-tuned as hyperparameters. Optimization parameters, including learning rate decay and momentum, are carefully optimized alongside these architectural choices. The models are trained using 80 percent of the data, with the remaining portion reserved for testing, with both datasets selected randomly. To identify the optimal model configuration, a random search approach (Bergstra and Bengio, 2012) is employed to optimize these hyperparameters effectively. Throughout the training process, the network minimizes a loss function derived from the partial likelihood function of the CPH model, enhancing the model's predictive efficacy in survival analysis. More detailed formulations and discussion on network architecture design followed in this analysis can be found in Kalatian and Farooq (2021).

Table 1 provides an overview of the model hyperparameters and corresponding performance metrics. The model architecture consists of two hidden layers, each comprising 30 nodes, with a dropout rate of 0.1 and batch normalization enabled. The learning rate is set at 0.0007, with a decay rate of 0.0003. Despite the robust architecture, the model's performance metrics indicate potential challenges. With a total of 10 covariates included, the model achieves a C-index of 0.628 on the validation set and 0.587 on the test set. Comparatively, the expanded covariate set may augment model complexity and the risk of overfitting relative to traditional CPH models. This complexity could impede the model's generalization ability, potentially leading to a lower Concordance Index.

Table 1: Deep CPH model parameters and performance

| Parameter | Value |
|---|---|
| Number of Hidden Layers | 2 |
| Number of Nodes in Hidden Layers | 30 |
| Dropout Rate | 0.1 |
| Batch Normalization | TRUE |
| Learning Rate | 0.0007 |
| Learning Rate Decay | 0.0003 |
| **Model Performance:** | |
| Number of covariates | 10 |
| C-index: validation set | 0.628 |
| C-index: test set | 0.587 |

In the traditional CPH model, the impact of each covariate on the log-partial hazard function is typically assessed through covariate coefficients. A higher coefficient indicates an increase in the log-partial hazard function, thereby influencing the hazard function accordingly. This implies that a positive coefficient for a covariate correlates with a greater likelihood of the event occurring sooner. However, traditional CPH models rely on z-scores and p-values to determine the significance of variables, which may not fully capture the complexities of nonlinear relationships among multiple variables in neural networks. To address this challenge, this analysis implements the SHAP (SHapley Additive exPlanations, Lundberg and Lee (2017)) technique. Originating from game theory principles, SHAP leverages Shapley Values to quantify the average marginal contribution of each feature to the model prediction for every instance. Unlike other interpretation methods, SHAP considers interactions among features, providing a comprehensive assessment of feature importance across all possible subsets. By utilizing Shapley values, the significance of features is evaluated based on their average absolute contributions across the entire dataset, thereby enhancing the interpretability of the neural network-based survival model.

*3.5. Multivariate regression for perceived mental workload*

The multivariate linear regression models are estimated to investigate the influential factors shaping perceived mental workload in takeover scenarios. The model elucidates the interplay among several independent variables ($X_1, X_2, ..., X_p$) and a dependent variable ($Y$). Its general formulation is expressed as:

$$Y = \beta_0 + \beta_1 X_1 + \beta_2 X_2 + ... + \beta_p X_p + \epsilon \tag{5}$$

Where:

$Y$ denotes the dependent variable.

$X_1, X_2, ..., X_p$ are the independent variables.

$\beta_0$ represents the intercept term.

$\beta_1, \beta_2, ..., \beta_p$ denote the coefficients linked with the independent variables.

$\epsilon$ stands for the error term, capturing unobserved factors impacting $Y$.

In multivariate linear regression, the aim is to estimate coefficients ($\beta_0, \beta_1, ..., \beta_p$) that minimize the sum of squared differences between observed and predicted values of $Y$. The models encompass both normalized continuous variables and binary categorical variables. For normalized continuous variables, linearity dictates that a one-unit change in the variable results in a proportional change in the dependent variable, as reflected by the coefficient. Conversely, dummy variables representing binary categories introduce an intercept shift. Each coefficient linked to these dummy variables indicates the difference in intercepts between their respective categories and the reference category. By incorporating categorical variables as dummy variables, the model captures the influence of these categories on the dependent variable, allowing for potential nonlinear relationships. This approach acknowledges that the impact of independent variables on perceived mental workload may deviate from linearity, enhancing the model's flexibility and realism.

**4. Data**

Controlled laboratory experiments were conducted using a virtual and immersive reality environment (VIRE, Farooq et al. (2018)) to examine the mandatory and discretionary takeover conditions. The takeover experience in VIRE settings is depicted in Figure 2.

*4.1. Study design*

The study adopts a mixed factorial design characterized by four controlled variables (listed in Table 2) with two levels each (2 × 2 × 2 × 2). Figure 2a and Figure 2b illustrate two example scenarios, providing a visual representation of how the underlying situations depict the control variables. The three distinct participant groups based on the types of driving control situations experience all 16 conditions. Every participant within a group encounters a random set of four conditions. It helps mitigate the order effects and manage participant burden and study duration. The design ensures equal repetition of scenarios within each group, minimizing potential biases. The underlying experimental design allows for efficient data collection while providing valuable insights into the diverse effects of the conditions on participants' responses within each of the three groups.

During a typical 30-minute experiment session, sociodemographic information, perceptual variables, and locus of control (Rotter, 1966) data were gathered through a pre-experiment ques-

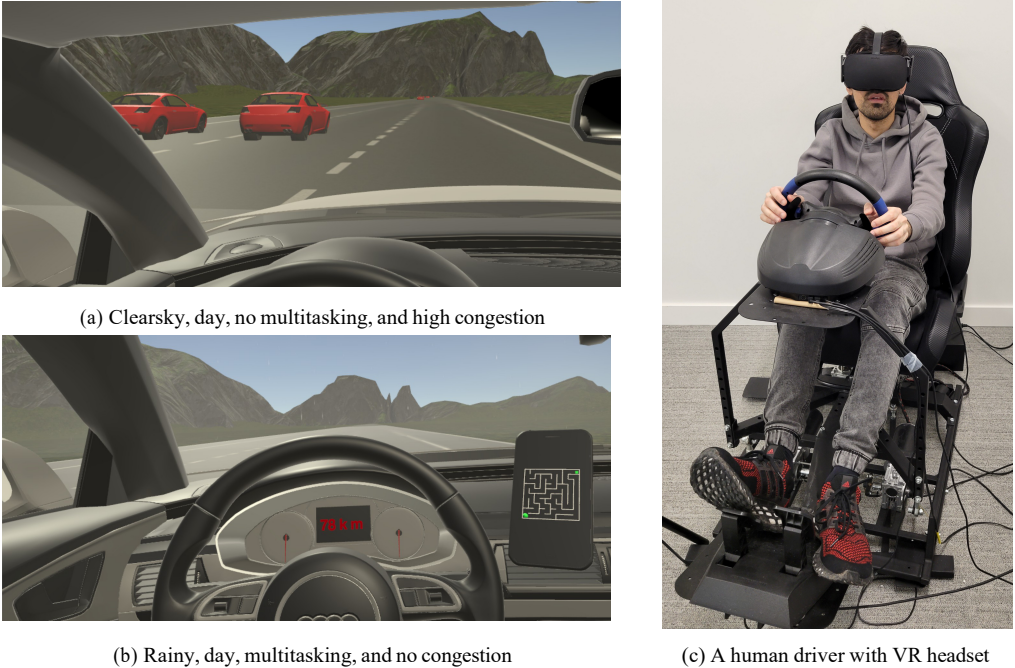

(a) Clearsky, day, no multitasking, and high congestion

(b) Rainy, day, multitasking, and no congestion

(c) A human driver with VR headset

Figure 2: Takeover experience in VIRE

Table 2: Description of variables and split of experiments

| variables | intervals | variable definition | observations (N = 304) | discretionary takeover (%) | mandatory takeover (%) |
|---|---|---|---|---|---|
| experiment group | mixed | mix of DTOR and MTOR | 104 | 50 | 50 |
| | DTOR | discretionary takeover scenarios only | 100 | 100 | - |
| | MTOR | mandatory takeover scenarios only | 100 | - | 100 |
| gender | male | - | 206 | 54 | 46 |
| | female | - | 98 | 40 | 60 |
| age | age_one | 18-24 years | 82 | 40 | 60 |
| | age_two | 25-29 years | 95 | 60 | 40 |
| | age_three | 30-39 years | 107 | 48 | 52 |
| | age_four | 40-65 years | 20 | 54 | 46 |
| job | job_1 | employed | 78 | 33 | 67 |
| | job_2 | student | 226 | 55 | 45 |
| education | edu_one | college/university | 56 | 44 | 56 |
| | edu_two | bachelors | 66 | 40 | 60 |
| | edu_three | masters | 114 | 62 | 38 |
| | edu_four | PhD | 68 | 44 | 56 |
| license | drive_one | G1 | 44 | 34 | 66 |
| | drive_two | G2 | 46 | 58 | 42 |
| | drive_three | G | 133 | 44 | 56 |
| | drive_four | international license | 81 | 66 | 34 |
| driving experience | drive_exp_one | <2 years | 66 | 59 | 41 |
| | drive_exp_two | 2-5 years | 82 | 56 | 44 |
| | drive_exp_three | 5-10 years | 68 | 44 | 56 |
| | drive_exp_four | >10 years | 88 | 43 | 57 |
| famAV | no | familiarity about AVs | 78 | 58 | 42 |
| | yes | | 226 | 48 | 52 |
| multi_tasking[1] | no | visual manual mounted secondary tasks | 152 | 50 | 50 |
| | yes | | 152 | 51 | 49 |
| weather[1] | clearsky/sunny | - | 150 | 50 | 50 |
| | rainy | - | 154 | 50 | 50 |
| lighting[1] | day | - | 150 | 52 | 48 |
| | night | - | 154 | 50 | 50 |
| traffic[1] | heavy_congestion | - | 152 | 50 | 50 |
| | light_congestion | - | 152 | 50 | 50 |

[1]controlled variables with two levels each (2 × 2 × 2 × 2)

tionnaire. Subsequently, the experiment transitioned to the simulator phase. After the training session, the vehicle motion data (10 frames per sec) were recorded in the actual tests. Following each scenario, participants were required to complete a self-reported NASA-TLX questionnaire to assess their perceived mental workload.

*4.2. Participants*

The data presented in Table 2 was gathered from 104 participants, resulting in at least 100 observations for each of the three experimental groups. The data collection process aimed to maintain an even split of discretionary and mandatory takeover conditions across all three groups. Approximately one-third of the dataset is sourced from our previous campaigns, Djavadian et al. (2019); Ansar et al. (2023a,b). Nonetheless, the experimental protocols and ethics remained consistent with the approach used in the present study.

The dataset represents a broad cross-section of the population, capturing a wide range of sociodemographic characteristics. For example, different age group composition highlights its inclusivity and the representation it offers. It encompasses participants with diverse educational backgrounds, Ontario driving license categories, and previous driving experience. Most participants, roughly three out of every four, demonstrated familiarity with AVs. Every scenario presented to the participants was constructed using four controlled variables, each having two levels. These variables included multitasking, weather conditions, lighting, and traffic congestion.

## 5. Result and Analysis

*5.1. Initial exploration of influencing factors*

In the preliminary analysis, we conducted a comprehensive exploration of influencing factors using two non-parametric statistical tests: the Kruskal-Wallis (KW) test and the Mann-Whitney U (MWU) test. Our focus centred on ten key factors crucial for shaping takeover performance and overall experience. These factors included perceived mental workload, locus of control, as well as kinematic indicators and stability indices. The analysis covered over 350 differences, comprising 6 group differences and 37 pairwise comparisons. These differences predominantly involved sociodemographic factors and within and between-group control variables. Notably, we identified 114 statistically significant differences, underscoring the considerable variability. These significant differences are displayed in the radial plot shown in Figure 3, offering insights into the magnitude of the observed distinctions.

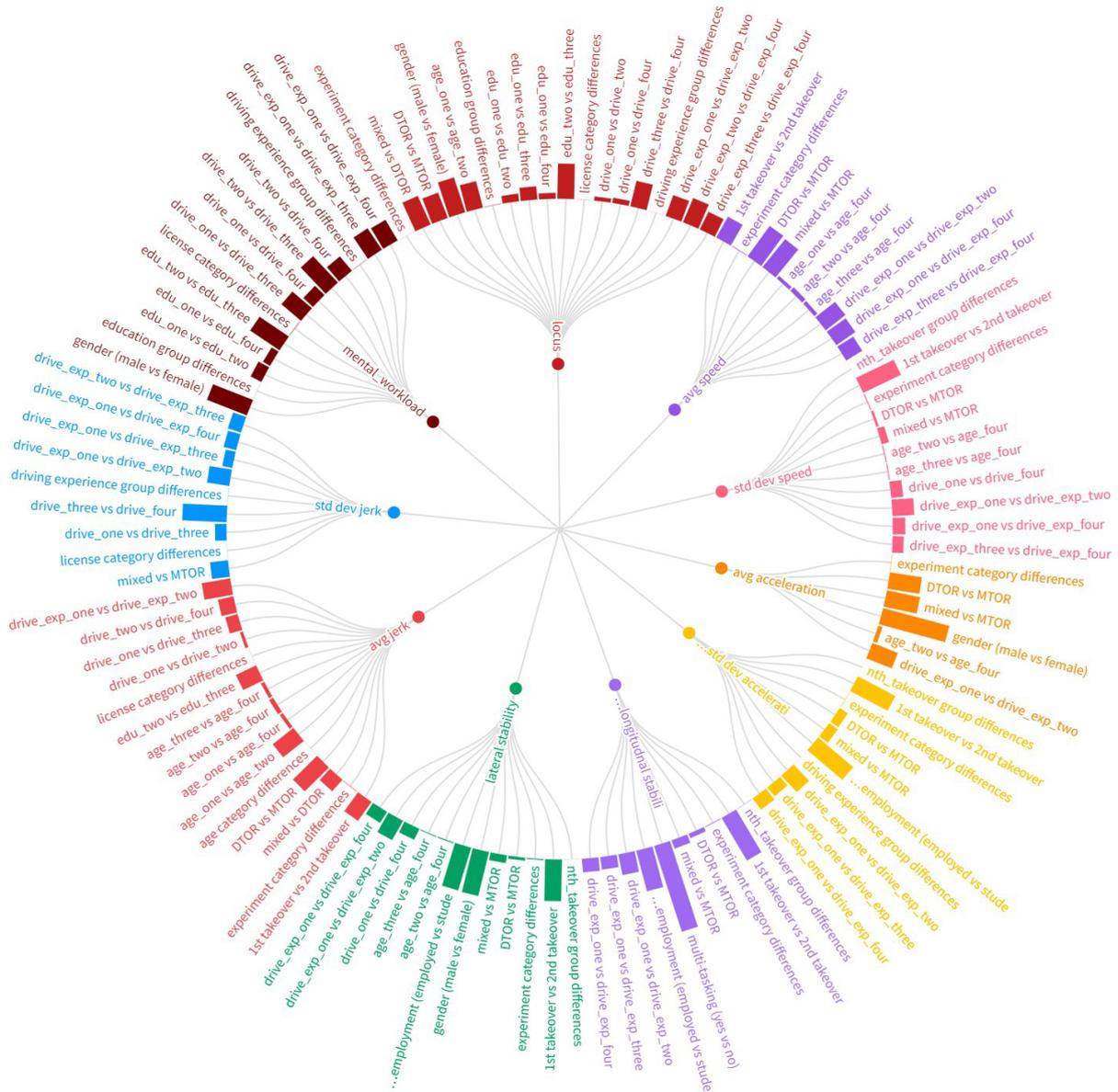

Figure 3: Radial plot for significant differences

A longer bar in the plot indicates higher test statistics. The bar represents KW statistics for group differences involving more than two categories and MWU statistics for pairwise comparisons as part of the post hoc analysis. For instance, ten bars of significant pairwise differences are depicted for the perceived 'mental workload.' It suggests that these specific pairwise comparisons exhibit a more substantial difference in mental workload, emphasizing the influential role of gen-der, education, and driving experience in shaping perceptions of mental workload during takeover scenarios. Similarly, the three KW statistics for perceived mental workload reveal notable differences. The test statistics observed for driving experience (i.e., "driving experience group differences"), license category, and education group differences demonstrate significant variations in mental workload levels among individuals during takeover scenarios.

The primary motivation behind this initial analysis was to identify and understand potential trends and significant influencing factors in driving takeover scenarios. By employing robust statistical tests, we aimed to discern patterns that could inform subsequent stages of our research. For instance, the significant covariates identified help in constructing the regression analysis for each key factor in the later analysis. This exploratory step lays the groundwork for a more in-depth investigation into the nuanced dynamics of driving takeover, providing essential insights into the factors contributing to the observed differences.

*5.2. Visualizing data distribution of influencing factors*

Examining the irregularities in kinematics and concerning key variables like lateral and longitudinal stability and the NASA TLX index among various experimental groups is essential. Violin plots help visualize data distribution, which uncovers the patterns which are not readily apparent in summary statistics. Figure 4 illustrates the distribution of kinematic parameters for each takeover type. On the y-axis, the normalized values of these kinematic parameters can be observed, while the x-axis represents the three distinct experiment groups. The width of the violin at any given point represents the data density, while the lines inside it represent the median, upper quartile, and lower quartile. For instance, the long dashed line corresponds to the median, representing the central value when the data is arranged from smallest to largest. The upper dashed line indicates the upper quartile, encompassing the top 25 percent of the data values, while the lower dashed line covers the lower quartile, representing the bottom 25 percent of the data values. The distinct shapes of the violin plots and the positions of the lines within them collectively signify variations in takeover behaviour among the experimental groups.

Kinematic parameters, including speed, acceleration, and jerk (the rate of change of acceleration), are depicted for each takeover scenario. The distribution differences in these variables are illustrated by the violin plots between "discretionary takeover" and "mandatory takeover" within the "mixed," "DTOR," and "MTOR" categories. A higher dispersion in mandatory takeover for average and standard deviation of speed is visible, reflecting the higher heterogeneous speed in the case of system-initiated takeovers. The median and high-density areas for average speed, acceleration, and jerk consistently indicate higher values during "discretionary takeover" than "mandatory takeover" in all three experimental groups. This implies that, on average, "discretionary takeover" results in greater speed, acceleration, and jerk in these experiments across all groups. Notably, the standard deviation for these variables is greater in "mandatory takeover," indicating more variability in responses compared to "discretionary takeover." However, the distribution difference of "discretionary takeover" in the "mixed" and "DTOR" categories is not as significant as that of "mandatory takeover" in the "mixed" and "MTOR" categories.

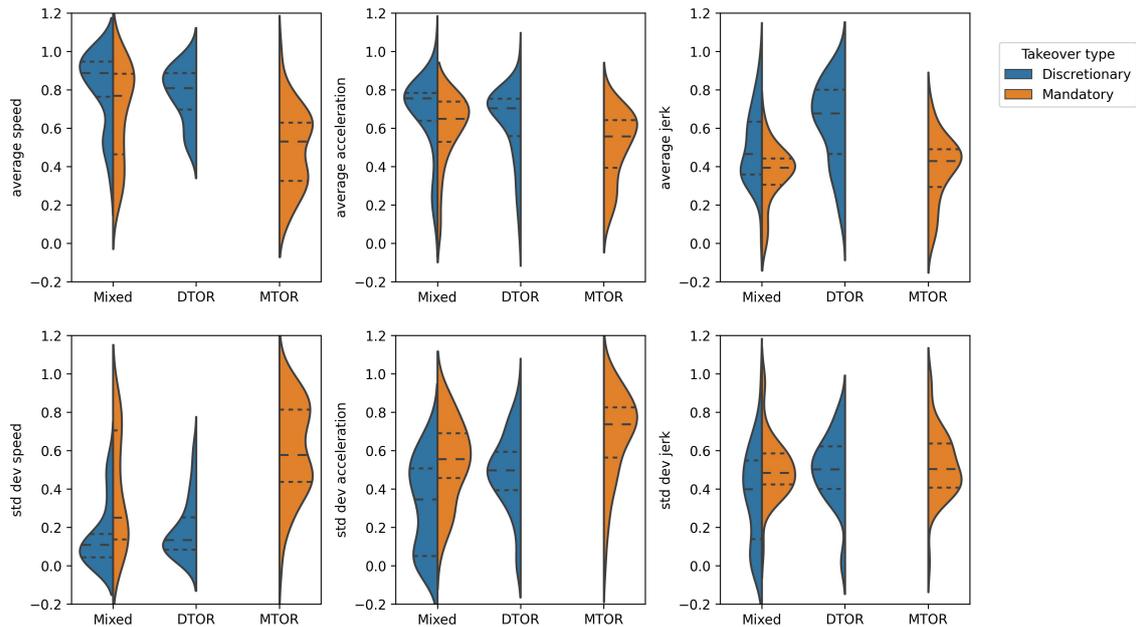

Figure 4: Violin plots for kinematic parameters

The plots presented in Figure 5 reveal heightened lateral and longitudinal instability and a more dispersed violin plot in mandatory takeover scenarios. This implies that drivers may encounter difficulties in adapting to the abrupt shift from automated to manual control. The substantial divergence in the distribution of lateral and longitudinal stability indicates variations in the complexity of takeover tasks among the experiment groups. Furthermore, the perceived men-tal workload during discretionary takeovers for the mixed experiment group is marginally lower than in mandatory takeovers. However, it exhibits greater dispersion in "DTOR" and an almost identical median when compared to "MTOR." This suggests that discretionary scenarios result in a more diverse range of subjective experiences among drivers.

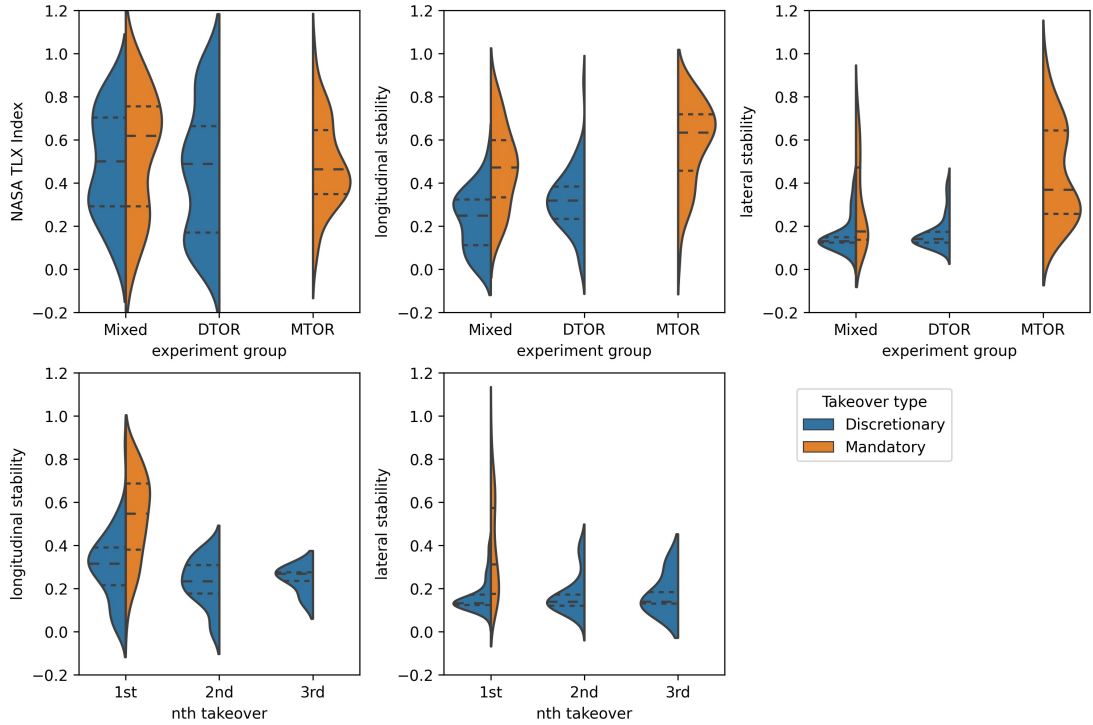

Figure 5: Violin plots for key variables distribution

In discretionary takeovers, where drivers independently initiate or terminate automation and encounter multiple takeover instances, we assess lateral and longitudinal stability for the first, second, and subsequent takeovers (third and beyond). By examining density distribution and dispersion, we observe (in the bottom-left plot) that longitudinal stability exhibits an inherent learning effect. Motivated by this finding, we delve deeper into analyzing the kinematics of the nth takeover to gain a more in-depth understanding of the evolving dynamics with each successive takeover event.

Table 3: kinematics in nth takeovers

| | | 1st takeover | | 2nd takeover | 3rd and subsequent takeovers | | | 1st takeover | | 2nd takeover | 3rd and subsequent takeovers |
|---|---|---|---|---|---|---|---|---|---|---|---|
| | | discretionary | mandatory | - | - | | | discretionary | mandatory | - | - |
| | count | 109 | 133 | 37 | 14 | | count | 109 | 133 | 37 | 14 |
| avg_speed (m/s) | mean | 23.66 | 21.04 | 23.46 | 23.72 | std_speed (m/s) | mean | 1.02 | 2.58 | 1.16 | 1.01 |
| | std | 2.72 | 3.35 | 2.81 | 3.22 | | std | 0.96 | 1.47 | 1.06 | 1.09 |
| | min | 16.28 | 14.95 | 15.71 | 18.62 | | min | 0.00 | 0.05 | 0.01 | 0.00 |
| | max | 26.92 | 26.86 | 26.91 | 26.91 | | max | 3.80 | 5.09 | 3.97 | 2.84 |
| avg_acc (m/s$^2$) | mean | -0.50 | -0.94 | -0.75 | -0.65 | std_acc (m/s$^2$) | mean | 1.22 | 1.80 | 1.19 | 1.14 |
| | std | 0.76 | 0.77 | 1.06 | 0.86 | | std | 0.73 | 0.65 | 0.63 | 0.66 |
| | min | -2.86 | -3.01 | -3.39 | -2.13 | | min | 0.02 | 0.06 | 0.02 | 0.01 |
| | max | 0.93 | -0.02 | 0.13 | 0.00 | | max | 2.71 | 2.88 | 2.16 | 1.71 |
| avg_jerk (m/s$^3$) | mean | 0.31 | 0.01 | 0.30 | 0.06 | std_jerk (m/s$^3$) | mean | 4.38 | 5.50 | 4.72 | 4.09 |
| | std | 0.31 | 0.23 | 0.32 | 0.28 | | std | 2.38 | 1.70 | 1.87 | 2.37 |
| | min | -0.30 | -0.51 | -0.46 | -0.38 | | min | 0.18 | 0.41 | 0.24 | 0.15 |
| | max | 0.89 | 0.54 | 0.75 | 0.34 | | max | 9.63 | 9.84 | 7.40 | 5.90 |
| long_stability | mean | 0.84 | 1.45 | 0.68 | 0.59 | lat_stability | mean | 0.10 | 0.17 | 0.10 | 0.10 |
| | std | 0.45 | 0.51 | 0.30 | 0.33 | | std | 0.04 | 0.08 | 0.05 | 0.04 |
| | min | 0.08 | 0.22 | 0.07 | 0.05 | | min | 0.05 | 0.07 | 0.04 | 0.07 |
| | max | 2.45 | 2.30 | 1.11 | 0.83 | | max | 0.22 | 0.33 | 0.22 | 0.17 |

## 5.3. Learning effect in nth takeovers

Table 3 focuses on the kinematic parameters, shedding light on the variations in speed, acceleration, jerk, and stability indices. The objective is to decipher the learning effect and evolving driving control during takeover scenarios. The trends across multiple discretionary takeovers suggest that drivers may exhibit a learning effect, gradually adapting and refining their takeover behaviour with each subsequent task. The standardized measures help quantify the consistency and stability of driving behaviour.

Examining the kinematic parameters of first, second, and subsequent takeovers in discretionary scenarios reveals meaningful variations. The initial takeover exhibits a higher average speed than the second, implying that drivers may approach the takeover more cautiously after the first experience. However, the third and subsequent takeovers show a slight increase in average speed (23.72 m/s), indicating a potential learning effect or increased confidence over time. A similar intuition regarding the learning effect is also justified in jerk (m/s³) values. Furthermore, lateral stability remains consistent across takeover attempts, while longitudinal stability shows marginal improvement with repeated experiences.

In comparing the initial discretionary takeover with its mandatory counterpart, distinctive differences emerge across various kinematic parameters. Throughout the first discretionary takeover, drivers demonstrated a higher average speed (23.66 m/s) than in the mandatory takeover (21.04 m/s). This pattern is further reflected in the standard deviation of speed, indicating greater variability in mandatory scenarios. Moreover, discretionary takeovers showcase a lower average acceleration (-0.50 m/s²) compared to mandatory takeovers (-0.94 m/s²), implying a more controlled and gradual acceleration profile in the former. Additionally, reinforcing this trend, discretionary takeovers exhibit higher longitudinal and lateral stability (indicated by lower indices) than mandatory takeovers (indicated by higher indices), suggesting a more stable control during the discretionary transitions.

*5.4. Survival models for Perception-Reaction Time (PRT)*

*5.4.1. Cox Proportional Hazard (CPH) model*

The study utilized the CPH model to investigate PRT for mandatory takeovers. Distinguishing between the events of safe takeover (when, PRT ≤ 3.55 sec) and unsafe takeover (when, PRT > 3.55 sec), the CPH analysis offers insights into the impact of various factors on the hazard ratio, representing the risk of experiencing unsafe events. The coefficients of covariates, hazard ratios, and their significance are summarized in Table 4. The data labels safe PRT events as 1 and unsafe events as 0.

Prior experience, as indicated by the variable 'control transition before,' exhibits a multiplicative effect on the likelihood of a safe transition. This suggests that individuals who have previously experienced a control transition have a lower hazard of unsafe PRT compared to those who have not encountered such transitions before. Additionally, a similar effect is observed with familiarity with AVs, indicating that individuals with prior familiarity with AVs have a lower hazard of unsafe PRT than those who lack such familiarity. Furthermore, drivers with 2 to 5 years of driving experience and those holding an international license were observed to be more cautious in their transitions. The coefficients and hazard ratios of 0.74 (2.09) and 0.57 (1.77) indicate that these individuals have a lower hazard of experiencing unsafe PRT, suggesting a safer transition behaviour than other groups.

A decrease in the locus of control index is linked to a reduced likelihood of encountering an unsafe takeover. Externalizers, identified as individuals with a higher locus index according to

Table 4: Multivariate proportional hazard model results

| Variable | Coefficient | Hazard ratio |
|---|---|---|
| locus of control index | -0.89 | 0.41 * |
| control transition before | 0.85 | 2.34 |
| familiarity about AVs | 0.39 | 1.48** |
| 2-5 years driving experience | 0.74 | 2.09 |
| international driving license | 0.57 | 1.77 * |
| heavy congestion | -0.51 | 0.6 |
| multitasking and rainy | -0.31 | 0.74** |
| **Performance indicators:** | | |
| Concordance Index: 0.69 | | |
| - log2(p) of ll-ratio test: 13.4 | | |

* Not statistically significant at 95% confidence level
** Not statistically significant at 90% confidence level

Ansar et al. (2023b), face a higher hazard of experiencing prolonged reaction times compared to internalizers. Likewise, adverse weather conditions (rainy), heavy traffic congestion, and engaging in multitasking notably compromise the safety of the transition. The risk of an unsafe takeover significantly increases under these circumstances.

In assessing the performance indicators, the underlying CPH model demonstrates a moderately good predictive ability with a C-index of 0.69. This suggests effective discrimination between individuals with varying hazard rates. Furthermore, the substantial significance of the model is underscored by the -log2(p) value of 13.4, signifying that the included predictors contribute valuable information, rendering the model statistically significant.

*5.4.2. Interpretable Deep CPH model*

A deep survival model, incorporating the SHAP explainability technique, has been integrated into the analysis alongside the traditional CPH model. This deep CPH model expands covariate estimation and possesses an enhanced capacity to capture nonlinear relationships, thereby providing a comprehensive understanding of survival dynamics by considering a broader array of influencing factors.

The analysis of SHAP mean and standard deviation (std) values in Table 5 offers valuable insights into the influence of each variable on the model's predictions and their consistency across instances. Positive SHAP mean values indicate a reduced hazard of unsafe PRT, while negative values suggest an increased hazard. For instance, the covariate "control transition before" exhibits a notable mean SHAP value of 0.708, indicating its significant contribution to the model's predictions and a reduced risk of unsafe PRT. This finding aligns with the intuitive understanding that prior experience leads to shorter reaction times, thereby enhancing overall safety. Similarly, "familiarity about AVs" presents a positive mean SHAP value of 0.636, suggesting its favourable impact on the model's predictions and indicating that greater familiarity with AVs is associated with quicker takeover responses. Notably, both covariates demonstrate effects consistent with the estimations of the CPH model, underscoring their robust influence in reducing the hazard of unsafe PRT.

Table 5: Interpretability of covariates in Deep CPH

| Variable | mean SHAP value | std SHAP value |
|---|---|---|
| control transition before | 0.708 | 0.201 |
| familiarity about AVs | 0.636 | 0.249 |
| age (25-29 yrs) | -0.081 | 0.196 |
| 2-5 years driving experience | -0.491 | 0.229 |
| 5-10 years driving experience | 1.145 | 0.288 |
| heavy congestion | 0.316 | 0.325 |
| night | 0.416 | 0.341 |
| multitasking | -0.425 | 0.385 |
| rainy | 0.722 | 0.402 |
| NASA TLX index (normalized) | 0.036 | 0.092 |

Variables related to driving experience, such as "2-5 years driving experience" and "5-10 years driving experience," showcase divergent effects, with the former contributing negatively (mean SHAP value of -0.491) and the latter positively (mean SHAP value of 1.145) to the model's predictions regarding unsafe PRT. However, the positive coefficient of 0.74 associated with "2-5 years driving experience" in the CPH model contradicts the negative contribution suggested by the Deep CPH model. This discrepancy underscores the complexities inherent in modelling PRT outcomes and suggests potential limitations in each approach. One plausible explanation for this inconsistency could be the Deep CPH model's capability to capture nonlinear relationships and interactions among covariates, uncovering nuances that traditional CPH models may overlook.

Assessing the consistency of SHAP values across instances for each covariate is essential. Uniformity is achieved when the absolute mean SHAP value surpasses the standard deviation, indicating a stable influence on the model's predictions across scenarios. Conversely, non-uniform

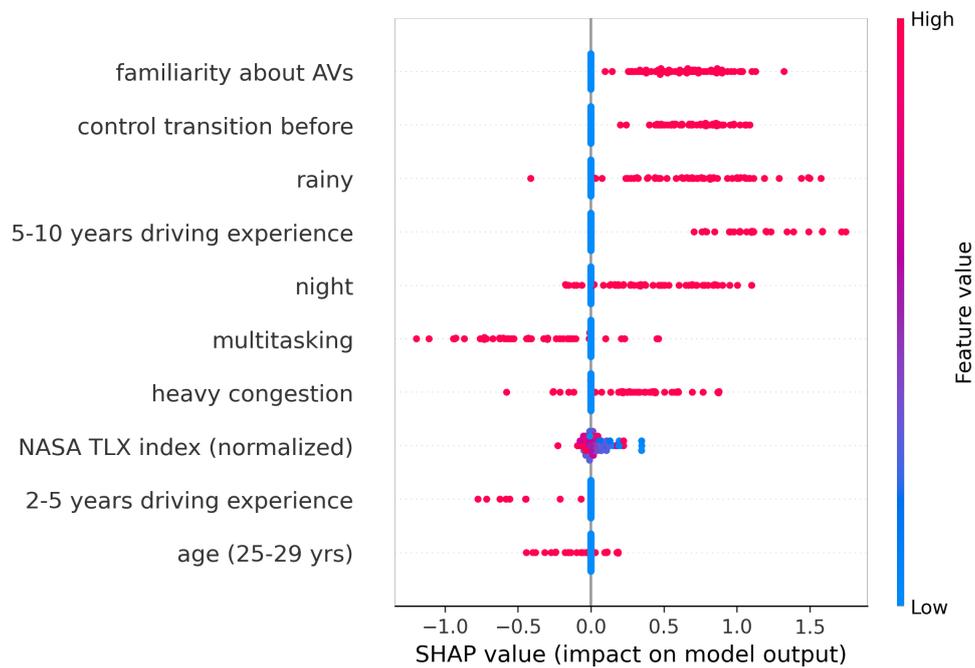

Figure 6: Plot of covariates impact based on SHAP Values

SHAP values suggest variability in the covariate's effects, requiring careful interpretation. For instance, covariates like "age (25-29 yrs)" with non-uniform SHAP values may exhibit diverse effects on PRT outcomes across instances, emphasizing the need for heterogeneity analysis. Additionally, as shown in Figure 6, the plot depicts SHAP values using dots, representing their impact on the model output. Red dots signify instances with values higher than the covariate's baseline, while blue dots denote instances with lower values. For binary covariates with a baseline of zero, instances with a zero value are excluded from impact calculations, as depicted by the blue dots along the central axis in the figure.

The dispersion of red dots in the summary plot indicates that driving during nighttime and in rainy weather conditions is associated with safer and faster takeover events, as evidenced by the positive SHAP values. This could be attributed to drivers being more cautious or attentive in these challenging driving conditions, thereby facilitating quicker responses during takeover events. Conversely, multitasking while driving, as indicated by its red dots on the negative side of the plot, elevates the risk of unsafe takeover events and extends the PRT, underscoring the detrimental impact of distractions on the safe operation of vehicles in automated driving modes.

Nevertheless, the relatively high variability across different instances highlights the necessity for targeted investigations into specific driving contexts.

*5.5. Multivariate linear regression for perceived mental workload*

This subsection presents the outcomes of three distinct multivariate linear regression models in Table 6. The models aim to elucidate the influential factors shaping mental workload in takeover scenarios. Our investigation encompasses a diverse range of predictors, from demographic attributes to driving-related experiences. By scrutinizing the model estimates, t-values, R-squared values, and log-likelihoods, we explore how specific factors manifest and interact within each experimental context, offering valuable insights into the drivers' cognitive demands during takeover scenarios.

The dependent variable in each model is the perceived mental workload, documented based on participant responses to the NASA TLX questionnaire after each experimental session. Comparing the three models, the model for 'discretionary takeovers' stands out with the highest R-squared value (0.56), suggesting that the selected variables in this model collectively explain 56 percent of the variance in mental workload. The models for 'all takeovers' and 'mandatory takeovers' exhibit lower R-squared values (0.29 and 0.25, respectively), indicating that their models explain a relatively smaller proportion of the variability in mental workload. The choice of influential variables also varies across the models, emphasizing the unique factors impacting mental workload in each takeover condition. Overall, these regression models provide valuable insights into the complex relationship between various factors and mental workload.

Several exploratory variables are found to be statistically significant on perceived mental workload. Notably, the significance of the nth takeover suggests that, beyond the initial transition, each subsequent takeover corresponds to an increase in mental workload. This aligns with the notion that repeated takeovers may introduce a cumulative cognitive burden on drivers. The lateral stability index emerges as a factor with varying impacts across the models. A higher lateral stability index signifies unstable lateral movements, contributing to an elevated cognitive load and increased perceived mental workload in discretionary takeovers. Furthermore, the locus of control index unveils an intriguing dimension. The significant impact indicates that individuals with a more external locus of control experience higher mental workload. This finding aligns with psychological theories suggesting that individuals with a more internal locus of control tend to feel more in command of their circumstances, potentially leading to a lower perceived mental workload.

Table 6: Linear regression estimation of perceived mental workload

| Parameters | All takeovers estimate | All takeovers t-value | Discretionary takeovers estimate | Discretionary takeovers t-value | Mandatory takeovers estimate | Mandatory takeovers t-value |
|---|---|---|---|---|---|---|
| constant (intercept) | 0.13 | 1.73 * | 0.192 | 1.98 | 0.27 | 4.36 |
| **Kinematic parameter** | | | | | | |
| lateral stability index | 0.23 | 3.68 | 0.304 | 2.45 | 0.106 | 1.31** |
| **Experiment related variable** | | | | | | |
| nth takeovers | 0.06 | 2.44 | 0.063 | 2.79 | - | - |
| **Perceptual variable** | | | | | | |
| locus of control index | 0.20 | 2.91 | 0.312 | 3.85 | - | - |
| **Attitudinal variables** | | | | | | |
| familiarity about AVs | - | - | -0.059 | -1.59** | 0.104 | 2.33 |
| rider AV shuttle | - | - | -0.153 | -3.56 | 0.248 | 3.67 |
| **Sociodemographics** | | | | | | |
| gender | -0.05 | -1.55** | -0.051 | -1.20** | - | - |
| age (30-39 yrs) | 0.077 | 2.688 | - | - | 0.069 | 1.71 * |
| age (40-60 yrs) | - | - | -0.425 | -2.38 | - | - |
| masters education | -0.08 | -3.259 | -0.109 | -3.277 | -0.093 | -2.07 |
| < 2years driving experience | 0.13 | 3.445 | 0.173 | 4.99 | - | - |
| G2 driving license | 0.08 | 1.82 * | - | - | - | - |
| **Environmental variables** | | | | | | |
| day | 0.08 | 2.56 | - | - | 0.099 | 2.04 |
| multitasking | 0.07 | 2.654 | 0.123 | 4.19 | - | - |
| night_congested | 0.14 | 3.81 | 0.054 | 1.68 * | 0.133 | 2.41 |
| **Performance indicators:** | | | | | | |
| Number of Parameters | 12 | | 12 | | 8 | |
| R-squared | 0.29 | | 0.56 | | 0.25 | |
| Adj. R-squared | 0.26 | | 0.53 | | 0.20 | |
| Log-Likelihood | 47.80 | | 58.40 | | 23.73 | |
| Akaike Information Criterion | -71.60 | | -92.79 | | -31.45 | |
| Bayesian Information Criterion | -28.69 | | -56.51 | | -9.71 | |

\* Not statistically significant at 95% confidence level
\*\* Not statistically significant at 90% confidence level

The attitudinal variables, particularly familiarity with AVs and experience riding in AV shuttles, exhibit distinctive behaviour in discretionary and mandatory takeover scenarios. In discretionary takeovers, these attitudinal factors demonstrate a mitigating effect on mental workload, as reflected by their negative coefficients in the specific model. Within the sociodemographic factors, male participants ("gender") exhibit lower cognitive load during discretionary takeovers. However, the significance of this variable is not strong. On the other hand, participants in the age range of 40 to 60 years experience a lower mental workload during discretionary takeovers, as indicated by the estimate and t-value of -0.425 (-2.38). This suggests that middle-aged participants handle discretionary tasks with less mental stress compared to their counterparts. The educational level of having a master's degree emerges as a significant factor in alleviating mental stress across both discretionary and mandatory takeover scenarios, indicating a potential association between higher education and cognitive efficiency during takeover events. Conversely, factors such as less driving experience and holding a lower category of driving license (G2, as per the Ministry of Ontario) contribute to heightened mental demand. Furthermore, environmental variables such as the time of day (day or night), engagement in multitasking, and exposure to traffic congestion exhibit varied yet consistently positive effects on the mental workload associated with takeover conditions.

## 6. Concluding Discussion

The study conducted a comprehensive analysis of cognitive demand and driving stability, focusing on both discretionary and mandatory takeover scenarios. We specifically examined the perception-reaction time associated with system-initiated mandatory takeovers and explored the learning effect observed in repeated driver-initiated discretionary takeovers. A total of 104 participants contributed to the study through a mixed factorial design. We leveraged advanced technologies like virtual reality and digital twins to create a controlled and realistic experimental setting. By utilizing this approach, the study overcame the limitations of traditional data collection methods. To the best of our knowledge, this research marks the first endeavour to thoroughly analyze and model the quality of both discretionary and mandatory takeovers, offering valuable insights into the unique characteristics of each task.

The data collection process played a crucial role in shaping our research methodology. The recruitment of participants for our 30-minute-long exploratory study posed several challenges. Firstly, the identification of individuals with diverse driving experiences and demographic backgrounds proved to be time-consuming. Moreover, the study's reliance on VR technology introduced an additional hurdle—potential VR sickness. Notably, a small subset of participants (less than 2%) reported experiencing dizziness during the VR experiment, a common challenge associated with VR technology. In strict adherence to ethical protocols, we promptly terminated the experiments for those individuals to prioritize their well-being. Subsequently, the data collected from these participants were excluded from our analysis to maintain the integrity and ethical standards of our study. Navigating through these challenges demanded meticulous planning and execution to ensure the validity and reliability of our study.

Incorporating four distinct analysis techniques, including statistical tests, violin plots for data distribution visualization, survival analysis, and linear regression, offers a comprehensive and multi-faceted approach to decoding takeover behaviour. Mann-Whitney U and Kruskal-Wallis tests provide valuable insights into potential differences in takeover behaviour across categorical variables, while violin plots enhance the understanding of data distribution, aiding in identifying patterns and outliers. The Deep CPH model extends the analysis to survival data, allowing for a nuanced exploration of time-to-event dynamics in takeover occurrences. Finally, linear regression facilitates the examination of linear relationships between variables. By employing this diverse set of analytical tools, the study benefits from a robust and thorough investigation, uncovering intricate patterns, relationships, and potential predictors in takeover behaviour that extend beyond the limitations of any single analysis method.

Insights gleaned from the data distribution and group differences depicted in radial and violin plots, especially concerning mental demand, highly contributed in the development of multivariate linear regression models. In general, the findings underscored the substantial impact of drivers' prior familiarity and experience with AVs. Beyond mitigating perceived mental workload, these factors positively affected perception-reaction times, potentially enhancing overall safety within automated driving environments. However, the unexpected findings regarding the lack of alleviation of perceived mental demand in mandatory takeover situations by these factors raise important concerns and underscore the need for further research. These results suggest that mandatory takeovers inherently demand a higher level of cognitive engagement and readiness from drivers compared to discretionary takeovers. In such critical moments, the necessity for immediate intervention may overshadow any potential benefits derived from familiarity or previous experience with automated driving systems. Moving forward, it would be valuable for future studies to delve deeper into whether both types of takeovers necessitate specialized training to effectively prepare drivers for the cognitive demands they entail.

In the analysis of perception-reaction time, our research underscores the critical importance of considering various environmental, perceptual, and sociodemographic factors in comprehensively understanding driver behaviour during takeover events. We observed that drivers may prioritize safety by exhibiting heightened vigilance and adopting a cautious approach, particularly in adverse weather conditions. However, their engagement in multitasking may elevate the likelihood of unsafe takeovers. Additionally, the analysis highlighted the challenges faced by relatively younger and less experienced drivers who may struggle to maintain situational awareness and make timely decisions during takeover events. This underscores the imperative of targeted training programs aimed at enhancing their readiness and responsiveness in effectively managing critical takeover situations.

The existing literature extensively covers mandatory takeovers, leaving a noticeable gap in understanding voluntary takeover performance. Human performance models in the context of driving behaviour also primarily focus on and revolve around mandatory tasks. However, our analysis found a more heterogeneous distribution of kinematic parameters and increased response variability in the discretionary takeover scenarios. This highlights the significance of examining vehicle design and ergonomics before deployment, especially in cases where takeovers are initiated voluntarily. As the landscape of automated driving evolves, acknowledging and addressing the unique challenges associated with discretionary takeovers becomes imperative for ensuring a seamless integration of automated systems and enhancing overall road safety. Further research in this domain is essential to inform the development of effective training programs and human-machine interfaces tailored to the intricacies of discretionary takeover situations.